\documentclass[aps,twocolumn]{revtex4}%
\pdfoutput=1
\usepackage{amsfonts}
\usepackage{amsmath}
\usepackage{amssymb}
\usepackage{graphicx}%
\begin{document}
\title{Singularities in Speckled Speckle: Screening }
\author{David A. Kessler}
\affiliation{Physics Department, Bar-Ilan University, Ramat-Gan IL52900, Israel}
\author{Isaac Freund}
\affiliation{Physics Department, Bar-Ilan University, Ramat-Gan IL52900, Israel}
\date{14 July 2008}

\begin{abstract}
We study screening of optical singularities in random optical fields with two
widely different length scales. \ We call the speckle patterns generated by
such fields \emph{speckled speckle}, because the major speckle spots in the
pattern are themselves highly speckled. \ We study combinations of fields
whose components exhibit short- and long-range correlations, and find unusual
forms of screening.\ \ 

\end{abstract}
\maketitle

\section{INTRODUCTION}

\begin{figure}
[pb]
\begin{center}
\includegraphics[
natheight=6.530200in,
natwidth=3.862300in,
height=2.9023in,
width=1.727in
]%
{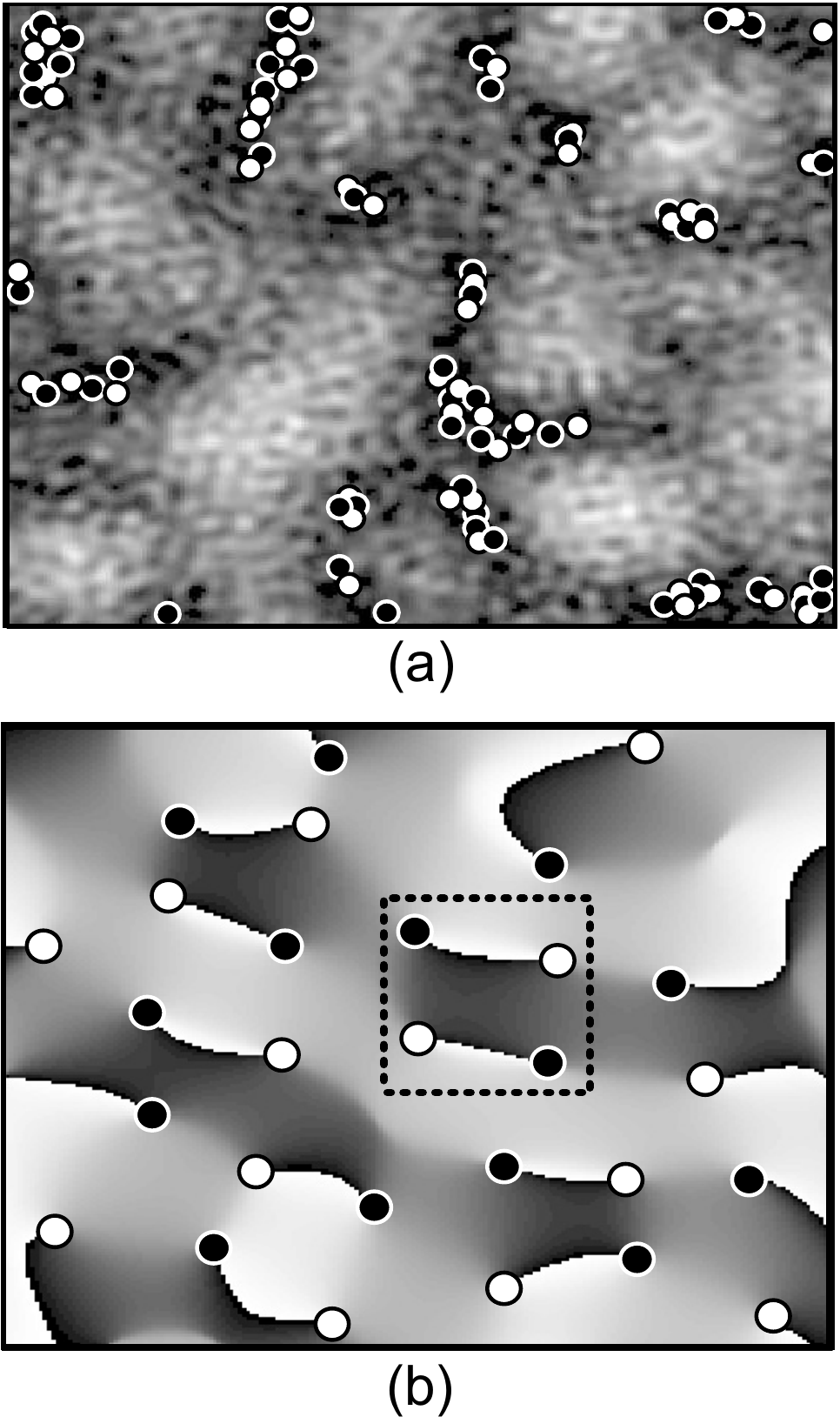}%
\caption{Vortex structures. \ Positive (negative) vortices are shown by white
(black) filled circles. \ (a) Speckled speckle. \ A random diffuser is
illuminated by two concentric disks of light, $a$ and $b$. \ The diameter of
disk $b$ is ten times the diameter of $a$; the total optical power in $a$,
however, is $10$ times that in $b$. \ Major (minor) speckle spots in the
speckled speckle field are due primarily to beam $a$ ($b$). \ Vortices of the
combined beam cluster in the dark regions between $a$ field speckle spots
because they require perfect destructive interference between the strong $a$
and weak $b$ fields. \ In normal speckle produced by a single disk vortices
tend to be uniformly distributed with only minor clustering. \ (b) Normal
speckle phase map produced by a random diffuser illuminated with a single
ring; the vortices tend to form a lattice with a square unit cell. }%
\label{Fig1}%
\end{center}
\end{figure}

Screening of charged topological singularities - vortices [1, 2 (Chap. 5),
$3$ (Sect. $4.8$)] in scalar fields, C points [$2,$ (Chaps. $12$ and $13$)] in
vector fields - has been extensively studied in random fields with a single
correlation length [$4-19$]; here we study screening of these singularities in
random fields with two widely different correlation lengths. \ We call such
fields \textquotedblleft speckled speckle\textquotedblright\ because, as
illustrated in Fig. \ref{Fig1}(a), the major speckle spots of the field are
themselves highly speckled. \ Speckled speckle fields can be generated by
illuminating a random diffuser with two concentric, overlapping beams: one,
the $a$ beam, is tightly focused and intense, the other, the $b$ beam, is
weak and diffuse.

The statistical properties of speckled speckle can be highly anomalous, with
relative number densities of critical points (vortices, C points, extrema, and
umbilic points) differing from normal \ speckle values by orders of magnitude
[$20,21$]. The spatial arrangement of vortices and C points is also anomalous,
with these singularities forming dense clusters of a kind not found in normal
speckle fields, Fig. \ref{Fig1}(a) [$20,21$].

Screening can be either short- or long-ranged. \ Nonsingular random sources
produce random fields that exhibit short range screening [$4-19$]. \ In such
systems positive (negative) topological charges are surrounded by a local net
excess of negative (positive) charge, leading to charge neutrality within a
characteristic distance, the screening length, that can be less than the
average separation between charges [$19$].

Singular sources, such as a ring of finite radius but zero width [$8$],
produce random fields that exhibits long-range screening [$8,16,19$]. \ The
singularities in the field produced by a ring form a quasi-lattice in which
positive/negative singularities occupy alternate corners of a square cell,
Fig. \ref{Fig1}(b). \ Local defects in the lattice destroy the local charge
neutrality that would produce short range screening, and screening sets in
only asymptotically.

Thus, both short- and long-range screening depend upon the spatial
arrangements of the charges. \ These arrangements are anomalous in speckled
speckle, which can therefore be expected to exhibit unusual forms of screening.

A major observable consequence of screening is strong damping of fluctuations
of the topological charge $Q$. \ These fluctuations are characterized
by their variance $\left\langle Q^{2}\right\rangle $; the behavior of
$\left\langle Q^{2}\right\rangle $ for speckled speckle is our main concern here.

The plan of this paper is as follows. \ In Section II we discuss the charge
variance in a bounded region, in Section III we review $\left\langle
Q^{2}\right\rangle $ in normal speckle for fields with short- and with
long-range correlations, and in Sections IV-VII we present results for
$\left\langle Q^{2}\right\rangle $ for four qualitatively different forms of
speckled speckle. \ We briefly summarize our main findings in the concluding
Section VIII.%

\section{CHARGE VARIANCE IN SHORT- AND LONG-RANGE SCREENING}

We assume isotropy, circular Gaussian statistics [$3$ (Chap 2), $22$ (Chap.
$2$)], and a circular region of radius $R$. \ The charge variance $\left\langle
Q^{2}\right\rangle $ in this region is related to the autocorrelation function
of the field $W(r)$ by [$19$ (Eq. 39)],%
\begin{equation}
\langle Q^{2}\rangle=\frac{1}{2\pi}\int_{0}^{2R}\sqrt{4R^{2}-r^{2}}%
\frac{(W^{\prime}(r))^{2}}{1-W^{2}(r)}dr. \label{Q2David}%
\end{equation}
where $W^{\prime}(r)=dW(r)/dr$. \ The number density of charges $\eta$ is
[$23$]%
\begin{equation}
\eta=-\frac{W^{\prime\prime}(0)}{2\pi}. \label{eta}%
\end{equation}

For the short range correlations produced by extended sources, $W^{\prime
}(r)  $ decays rapidly with $r$, and in the limit of large $R$,%
\begin{equation}
\langle Q^{2}\rangle\approx\frac{R}{\pi}\int_{0}^{\infty}\frac{(W^{\prime
}(r))^{2}}{1-W^{2}(r)}dr. \label{Q2short}%
\end{equation}
Thus, for short-range screening $\left\langle Q^{2}\right\rangle $ grows with
the perimeter, i.e. $\left\langle Q^{2}\right\rangle \sim R\sim\sqrt{N}$, in
contrast to the case of no screening, where $\left\langle Q^{2}\right\rangle $
grows with the area, $\left\langle Q^{2}\right\rangle \sim R^{2}\sim N$
[$8,19$].

\begin{figure}
[pt]
\begin{center}
\includegraphics[
natheight=8.368800in,
natwidth=3.986800in,
height=5.1301in,
width=2.4587in
]%
{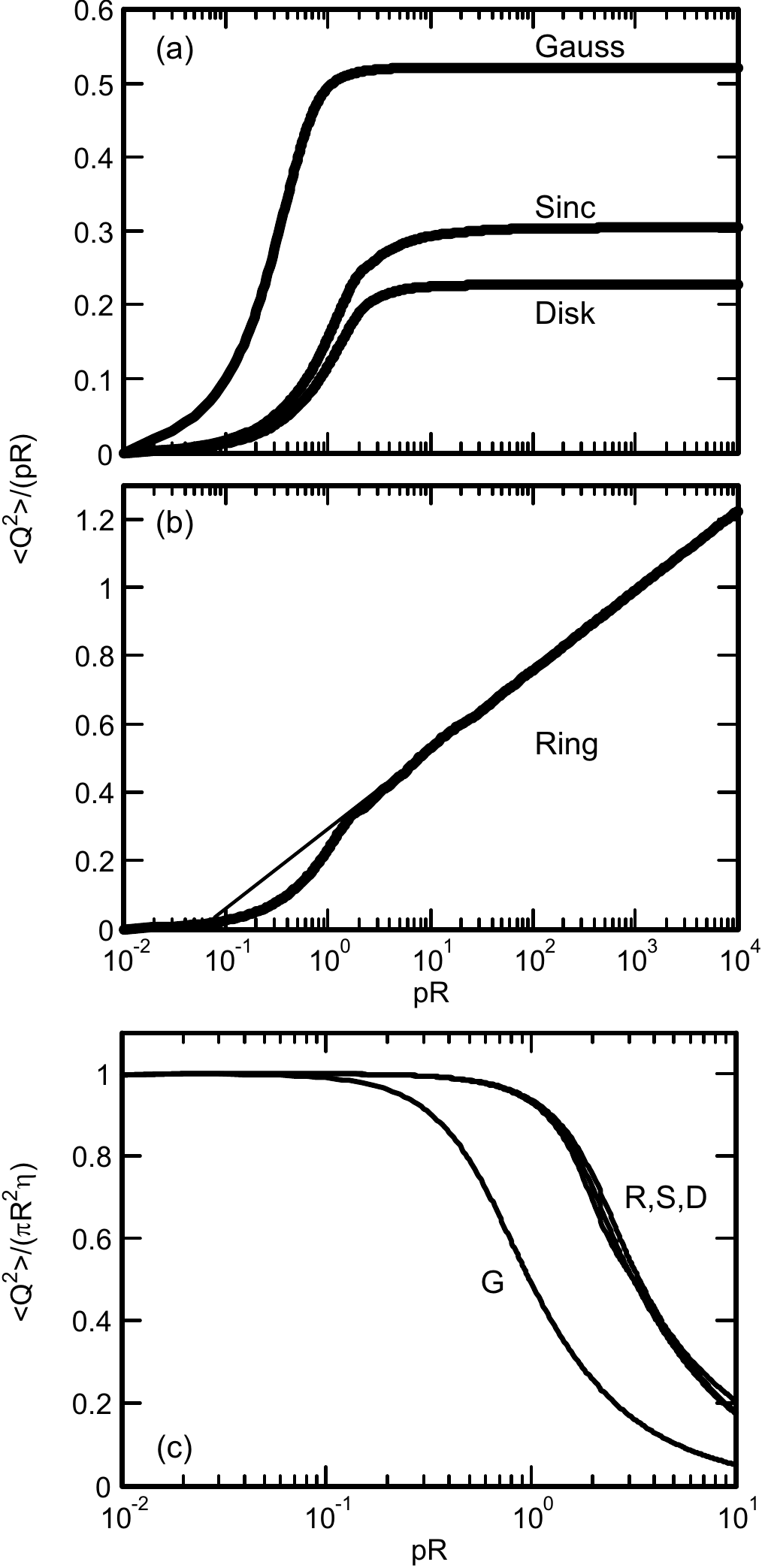}%
\caption{Charge variance $\left\langle Q^{2}\right\rangle $ for normal speckle
obtained from numerical integration of Eq. (\ref{Q2David}). \ (a) Short-range
screening. \ (b) Long-range screening. \ The dependence on the parameter $p$
in Eqs. (\ref{Gauss})-(\ref{Ring}) is here scaled out by plotting
$\left\langle Q^{2}\right\rangle /\left(  pR\right)  $ vs. $pR$. \ As can be
seen, for $pR>1$, the results in (a) quickly asymptote to the theoretical
values in Eqs. (\ref{LambdaScrnGauss})-(\ref{Q2pRSinc}), whereas the result in
(b) asymptotes to the theoretical form (thin straight line) in Eq.
(\ref{LongRangeQ2}). \ (c) $\left\langle Q^{2}\right\rangle $ for small $R$.
\ In all four cases (G - Gauss, D - Disk, S - Sinc, and R - Ring), the curves
approach the $R\rightarrow0$ limit given in Eq. (\ref{SmallRlimit}). }%
\label{Fig2}%
\end{center}
\end{figure}

For the long-range screening produced by a singular ring source of radius $p$
and zero width, the large $R$ limit is [$19$ (Eq. 48)],
\begin{subequations}
\label{LongRangeQ2}%
\begin{align}
\left\langle Q^{2}\right\rangle  &  \approx\dfrac{pR}{\pi^{2}}\left[
{\cal{K}}+\ln\left(  \rho R\right)  \right]  ,\label{LRQ2}\\
{\cal{K}}  &  =\pi{\cal{D}}+\gamma+5\ln2-3\approx2.81182,\label{DavidsK}\\
{\cal{D}}  &  =\int_{0}^{\infty}dx\,\frac{J_{0}^{2}(x)J_{1}^{2}(x)}%
{1-J_{0}^{2}(x)}\approx0.563047, \label{DavidsD}%
\end{align}
where $\gamma\approx0.577216$ is Euler's constant. \ Thus, long-range
screening yields a charge variance that grows asymptotically as $R\ln R$; this
growth rate is significantly faster than the short-range growth rate
proportional to $R$, but is very much slower than the unscreened rate
proportional to $R^{2}$. \ For, say, a large region that contains $10^{4}$
charges, short-range (long-range) screening damps out charge fluctuations
relative to no screening by a factor of $\sim100$ ($\sim22$).

\section{CHARGE VARIANCE IN NORMAL SPECKLE}

We review here the charge variance in normal speckle produced by sources with
a single characteristic length scale. \ In later sections we build our
composite sources with their two different length scales from binary
combinations of these single sources, and compare composite-source charge
variances with single-source variances.

\subsection{Source Distributions and Autocorrelation Functions}

Listed below are the source distributions $S(u)  $, where $u$
measures radial displacements in the source plane, the total optical power  in each
source, $P$, and the autocorrelation functions $W(r)  $ of
the speckle field. \ $S(u)$ and $W(r)$ are
related by the VanCittert-Zernike theorem [$3$ (Sect. 4), $22$ (Sect. 5.6)].
\ We study four fields with autocorrelation functions that decay at different rates.

(\emph{i}) A Gaussian, superscript (G), of $1/e$ width $p$, and intensity
(optical power/unit area) $I^{\left(  \text{G}\right)  }$ at the peak;
henceforth the \textquotedblleft Gaussian\textquotedblright,
\end{subequations}
\begin{subequations}
\label{Gauss}%
\begin{align}
S^{\left(  \text{G}\right)  }(u)  &  =I^{\left(  \text{G}\right)  }\exp\left(
-\left[  u/\left(  2p\right)  \right]  ^{2}\right)  ,\label{Gauss_S}\\
P^{\left(  \text{G}\right)  }  &  =4\pi p^{2}I^{\left(  \text{G}\right)
},\label{Gauss_P}\\
W^{(\text{G})}(r)   &  =\exp\left(  - p^2r^2\right)  . \label{Gauss_omega}%
\end{align}

(\emph{ii}) A uniform disk, superscript (D), of radius $p$, and uniform
intensity $I^{\left(  \text{D}\right)  }$; henceforth the \textquotedblleft
Disk\textquotedblright,
\end{subequations}
\begin{subequations}
\label{Disk}%
\begin{align}
S^{\left(  \text{D}\right)  }(u)  &  =I^{\left(  \text{D}\right)  }%
\Theta\left(  u-p\right)  ,\label{Disk_S}\\
P^{\left(  \text{D}\right)  }  &  =\pi p^{2}I^{\left(  \text{D}\right)
},\label{Disk_P}\\
W^{\left(  \text{D}\right)  }(r)   &  =2J_{1}\left(  pr\right)
/\left(pr\right)  . \label{Disk_omega}%
\end{align}
$\Theta\left(  x\right)  $ is the Heaviside step function defined by
$\Theta\left(  x\leq0\right)  =0$, $\Theta\left(  x>0\right)  =1$.

(\emph{iii}) A nonuniform (inverse square root) disk, superscript (S), of
radius $p$, with intensity $I^{\left(  \text{S}\right)  }$ at the disk center,%
\end{subequations}
\begin{subequations}
\begin{align}
S^{\left(  \text{S}\right)  }\left(  u\right)   &  =\frac{I^{\left(
\text{S}\right)  }}{\sqrt{1-\left(  u/p\right)  ^{2}}}\Theta\left(
u-p\right)  ,\label{Sinc_S}\\
P^{\left(  \text{S}\right)  }  &  =2\pi p^{2}I^{\left(  \text{S}\right)
},\label{Sinc_P}\\
W^{\left(  \text{S}\right)  }(r)   &  =I^{\left(  \text{S}%
\right)  }\text{sinc}\left(  pr\right)  , \label{Sinc_omega}%
\end{align}
where, sinc$\left(  x\right)  \equiv\sin\left(  x\right)  /x$. \ In what
follows we refer to this source as the \textquotedblleft
Sinc\textquotedblright.

(\emph{iv}) A singular ring, superscript (R), of radius $p$, which we write
as
\end{subequations}
\begin{subequations}
\label{Ring}%
\begin{align}
S^{\left(  \text{R}\right)  }(u)  &  =I^{\left(  \text{R}\right)  }%
\varepsilon\delta\left(  u-p\right)  ,\label{Ring_S}\\
P^{\left(  \text{R}\right)  }  &  =2\pi p\varepsilon I^{\left(  \text{R}%
\right)  },\label{Ring_P}\\
W^{\left(  \text{R}\right)  }(r)   &  =J_{0}(pr)  ,
\label{Ring_Omega}%
\end{align}
where $\delta\left(  x\right)  $ is the Dirac delta function. \ In what
follows we refer to this source as the \textquotedblleft
Ring\textquotedblright.

$S^{\left(  \text{R}\right)  }(u)$ is the limit of a finite width annulus
$s^{\left(  \text{R}\right)  }\left(  u\right)  $ of mean radius $p$ and width
$\varepsilon$,
\end{subequations}
\begin{subequations}
\label{Annulus}%
\begin{gather}
s\left(  u\right)  =\Theta\left(  u-p-\varepsilon/2\right)  -\Theta\left(
u-p+\varepsilon/2\right)  ,\label{Annulus_s}\\
\lim_{\varepsilon\rightarrow0}\left[  s\left(  u\right)  /\varepsilon\right]
=\delta\left(  u-p\right)  . \label{Annulus_S}%
\end{gather}
$I^{\left(  \text{R}\right)  }$ is therefore the uniform intensity in the
annulus. \ In the limit $\varepsilon\rightarrow0$, $I^{\left(  \text{R}%
\right)  }$ diverges, $P^{\left(  \text{R}\right)  }$ in Eq. (\ref{Ring_P}),
however, is assumed to remain finite.

$W$ and $\left\langle Q^{2}\right\rangle $ for the finite width annulus
$\left(  \varepsilon>0\right)  $ is discussed in [$19$ (Sect. 5)], where it is
shown that over the region $\varepsilon r<1$ the experimentally attainable
annulus is an excellent approximation to the theoretical singular ring.

\subsection{Charge Variance}

In Fig. \ref{Fig2} we plot $\left\langle Q^{2}\right\rangle /\left(
pR\right)  $ vs. $pR$ for the above four sources. \ For the Gaussian (G), Disk
(D), and Sinc (S), screening is short-ranged, and the large $R$ limit of
$\left\langle Q^{2}\right\rangle $ is given in Eq. (\ref{Q2short}).

For the Gaussian, Eq. (\ref{Q2short}) can be evaluated analytically, yielding%
\end{subequations}
\begin{equation}
\left\langle Q^{2}\right\rangle _{Gauss}\approx \tfrac{1}{4}\sqrt{2/\pi}\zeta\left(
3/2\right)  pR=0.521093pR, \label{LambdaScrnGauss}%
\end{equation}
with $\zeta\left(  x\right)  $ is the Riemann zeta function, whereas for the
Disk (D) and Sinc (S), Eq. (\ref{Q2short}) is evaluated numerically, yielding%
\begin{equation}
\left\langle Q^{2}\right\rangle _{Disk}\approx0.227210pR, \label{Q2pRDisk}%
\end{equation}
and%
\begin{equation}
\left\langle Q^{2}\right\rangle _{\operatorname{Si}nc}\approx0.305898pR.
\label{Q2pRSinc}%
\end{equation}

For the Ring (R) screening is long-ranged, and the large $R$ limit of
$\left\langle Q^{2}\right\rangle /\left(  pR\right)  $ is given in Eq.
(\ref{LongRangeQ2}).

For very small $R$ for all sources [$19$ (Eq. 82)],%
\begin{equation}
\left\langle Q^{2}\right\rangle _{R\rightarrow0}\approx\eta\pi R^{2}=N,
\label{SmallRlimit}%
\end{equation}
i.e. there is no screening; the reason is that for a sufficiently small area
the probability of finding the required screening charges within the area is
vanishingly small. \ This result is illustrated in Fig. \ref{Fig2}(c) for all four sources.%

\section{COMPOSITE SOURCES AND THEIR AUTOCORRELATION FUNCTIONS}

For scalar (single component) fields the $a$ and $b$ beams have the same, say,
linear polarization, and the singularities whose screening is of interest here
are the phase vortices [$1-19$]. \ For vector (two component) fields the $a$
and $b$ beams have orthogonal linear polarizations, and the relevant
singularities that screen each other are either right- or left-handed C points
[$2$]: right-handed C points do not screen left-handed ones, and vice versa.

We write our composite source as%
\begin{equation}
S^{\left(  \text{T}_{a}\text{T}_{b}\right)  }\left(  u\right)  =S_{a}^{\left(
\text{T}\right)  }\left(  u\right)  +S_{b}^{\left(  \text{T}%
\right)  }\left(  u\right)  , \label{CompositeS}%
\end{equation}
where the source type specifier $\left(  \text{T}_{a}\text{T}_{b}\right)  $ is
a binary combination of T$_{a,b}=$ G, D ,S, R. \ From the VanCittert-Zernike
theorem, the corresponding autocorrelation function is%
\begin{equation}
W^{\left(  \text{T}_{a}\text{T}_{b}\right)  }(r)  =\frac
{W_a^{\left(  \text{T}\right)  }\left(r\right)  +K^{\left(  \text{T}_a%
\text{T}_b\right)  }W_b^{\left(  \text{T}\right)  }\left(r\right)
}{1+K^{\left(  \text{T}_{a}\text{T}_{b}\right)  }}, \label{CompositeW}%
\end{equation}
where the dimensionless constant
\begin{equation}
K^{\left(  \text{T}_{a}\text{T}_{b}\right)  }=P_{b}^{\left(  \text{T}%
\right)  }/P_{a}^{\left(  \text{T}\right)  }. \label{K}%
\end{equation}
$S_{a,b}^{\left(  \text{T}\right)  }(u)  $, $P_{a,b}^{\left(
\text{T}\right)  }$, and $W_{a,b}^{\left(  \text{T}\right)  }\left(
r\right)  $, are listed in Eqs. (\ref{Gauss})-(\ref{Ring}), with $p=a,b$ as appropriate.

Similarly, the number density of singularities is for composite scalar fields,%
\begin{equation}
\eta^{\left(  \text{T}_{a}\text{T}_{b}\right)  }=-\frac{1}{2\pi}%
\frac{W_a^{\prime\prime\left(  \text{T}\right)  }(0)
+K^{\left(  \text{T}_{a}\text{T}_{b}\right)  }W_b^{\prime\prime\left(
\text{T}\right)  }(0)  }{1+K^{\left(  \text{T}_{a}%
\text{T}_{b}\right)  }}. \label{CompositeEta}%
\end{equation}
This is also the number density of right- and of left-handed C points.

As a specific example, for the intense, tightly focused $a$ beam a Gaussian
(G), and the diffuse weak $b$ beam a Sinc (S),
\begin{subequations}
\label{CompositeExample}%
\begin{align}
S_{a}^{\left(  \text{G}\right)  }(u) &  =I_{a}^{\left(  \text{G}\right)  }%
\exp\left(  -\left[  u/\left(  2a\right)  \right]  ^{2}\right)
,\label{CompositeExample_Sa}\\
S_{b}^{\left(  \text{S}\right)  }\left(  u\right)   &  =\frac{I_{b}^{\left(
\text{S}\right)  }}{\sqrt{1-\left(  u/b\right)  ^{2}}}\Theta\left(
u-b\right)  ,\label{CompositeExample_Sb}\\
W^{\left(  \text{GS}\right)  }\left(  r\right)   &  =\frac{\exp\left(
-ar\right)  +K^{\left(  \text{GS}\right)  }\text{sinc}\left(  br\right)
}{1+K^{\left(  \text{GS}\right)  }},\label{CompositeExample_W}\\
K^{\left(  \text{GS}\right)  } &  =\left[  2b^{2}I_{b}^{\left(  \text{S}%
\right)  }\right]  /\left[  a^{2}I_{a}^{\left(  \text{G}\right)  }\right]
,\label{CompositeExample_K}\\
\eta^{\left(  \text{GS}\right)  } &  =\frac{2a^{2}+K^{\left(  \text{GS}%
\right)  }b^{2}/3}{1+K^{\left(  \text{GS}\right)  }}%
,\label{CompositeExample_Eta}%
\end{align}

In the composite-source examples that follow we usually take $K\sim0.01$ and
$b/a=100$. \ There are two reasons for these choices: (\emph{i}) small $K$
together with large $b/a$ produces results that vividly illustrate the unusual
screening properties of speckled speckle, and (\emph{ii}) this combination of
parameters permits a significant degree of analysis. \ We consider that these
parameters, which are convenient for the theoretician, to be experimentally
possible; admittedly, they may be difficult to achieve in practice.

With the above choice of parameters, the $R$ dependence of $\left\langle
Q^{2}\right\rangle $ separates into three distinct regions:

\bigskip

I. $R<1/b$.

In this region Eq. (\ref{SmallRlimit}) holds for all composite sources with
$\eta$ equal to the number density of $b$ field charges%
\end{subequations}
\begin{equation}
\eta\approx\eta_{b}\approx-\frac{K}{2\pi}W_b^{\prime\prime\left(  \text{T}\right)
}(0)  .\label{Omegab}%
\end{equation}
The reason is that for sufficiently small $R$ the probability of finding an
$a$ beam charge in the area is negligible; only $b$ beam charges are present,
so only these charges contribute to $\left\langle Q^{2}\right\rangle $. \ This
result is verified by direct comparison (not shown) with the exact result in
Eq. (\ref{Q2David}).

\bigskip

II. $1/b<R<1/a$.

In this region, $a$ in Eq. (\ref{CompositeW}) may be set equal to zero,
because for $ar<1$, $W_a(r)  \approx W_a(0)  =1$, and
$W_a^{\prime}(r)  \approx W_a^{\prime}(0)  =0$,
independent of the $a$ beam type - G, D, S, or R. \ As will become apparent, these
approximations yield good agreement with the exact result in Eq.
(\ref{Q2David}).

\bigskip

III. $R>1/a.$

In this region there is no accurate approximation that is applicable, however,
as discussed below, approximations good to $\sim10\%$ are available.

Below we discuss screening in composite beams for the four qualitatively
different combinations of $a$ and $b$ beams in which the individual beams
exhibit either short- or long-range correlations.

\section{BOTH BEAM $a$ AND BEAM $b$ EXHIBIT SHORT-RANGE SCREENING}

We start with region II, $1/b<R<1/a$. \ Here the area contains many $b$
charges but practically no $a$ charges. \ We denote the $b$ charge
contribution to $\left\langle Q^{2}\right\rangle $ by $\left\langle
Q^{2}\right\rangle _{b}$, where%
\begin{equation}
\left\langle Q^{2}\right\rangle _{b}\approx\frac{R}{2\pi}K^{\left(
\text{T}_{a}\text{T}_{b}\right)  }\int_{0}^{\infty}\frac{\left[  dW_b^{\left(
\text{T}\right)  }(r)  /dr\right]  ^{2}}{1-W_b^{\left(
\text{T}\right)  }(r)  }dr. \label{SSQ2b}%
\end{equation}
In obtaining this result we make use of the fact that $K\ll1$ and $\left[
KW_b^{\left(  \text{T}\right)  }\right]  ^{2}\ll W_b^{\left(  \text{T}%
\right)  }$.

For $b$ a Gaussian (G), Eq. (\ref{SSQ2b}) can be evaluated analytically, and
we have,%
\begin{align}
\left\langle Q^{2}\right\rangle _{b}^{\left(  G\right)  }  &  \approx\frac
{K^{\left(  \text{T}_{a}\text{G}\right)  }b\left[  \zeta\left(  \frac{3}%
{2}\right)  -1\right]  }{2\sqrt{\pi}}R,\label{SSQ2bG}\\
&  =0.454843K^{\left(  \text{T}_{a}\text{G}\right)  }bR.\nonumber
\end{align}

For $b$ a Disk or a Sinc, Eq. (\ref{SSQ2b}) is evaluated numerically, and we
have for the Disk (D),%
\begin{equation}
\left\langle Q^{2}\right\rangle _{b}^{\left(  D\right)  }\approx0.187153K^{\left(
\text{T}_{a}\text{D}\right)  }bR, \label{SSQ2bD}%
\end{equation}
and for the Sinc (S),%
\begin{equation}
\left\langle Q^{2}\right\rangle _{b}^{\left(  S\right)  }\approx0.238531K^{\left(
\text{T}_{a}\text{S}\right)  }bR. \label{SSQ2bS}%
\end{equation}

In region III, $R>1/a$, we assume that the $b$ charges continue to contribute
$\left\langle Q^{2}\right\rangle _{b}$ to $\left\langle Q^{2}\right\rangle $.
\ In addition, the area now includes many $a$ charges. \ We label the
contribution of these charges $\left\langle Q^{2}\right\rangle _{a}$, and
write%
\begin{equation}
\left\langle Q^{2}\right\rangle \sim\left\langle Q^{2}\right\rangle
_{a}+\left\langle Q^{2}\right\rangle _{b}, \label{SSQaPlusQb}%
\end{equation}
i.e., we neglect interactions between the $a$ and $b$ beams, and the
possibility that $a$ charges can screen $b$ charges, and vice versa. \ Within
the framework of this approximation we write,%
\begin{equation}
\left\langle Q^{2}\right\rangle _{a}^{\left(  G\right)  }=\left\langle
Q^{2}\right\rangle _{Gauss},\text{ }\left[  \text{Eq.(\ref{LambdaScrnGauss}%
)}\right]  \label{SSQ2aG}%
\end{equation}%
\begin{equation}
\left\langle Q^{2}\right\rangle _{a}^{\left(  D\right)  }=\left\langle
Q^{2}\right\rangle _{Disk},\text{\ }\left[  \text{Eq.(\ref{Q2pRDisk})}\right]
\label{SSQ2aD}%
\end{equation}%
\begin{equation}
\left\langle Q^{2}\right\rangle _{a}^{\left(  S\right)  }=\left\langle
Q^{2}\right\rangle _{\operatorname{Sinc}},\text{ }\left[
\text{\ Eq.(\ref{Q2pRSinc})}\right]  \label{SSQ2aSinc}%
\end{equation}
with $p$ replaced by $a$.%

\begin{figure}
[pt]
\begin{center}
\includegraphics[
natheight=8.284900in,
natwidth=3.971200in,
height=4.9986in,
width=2.4102in
]%
{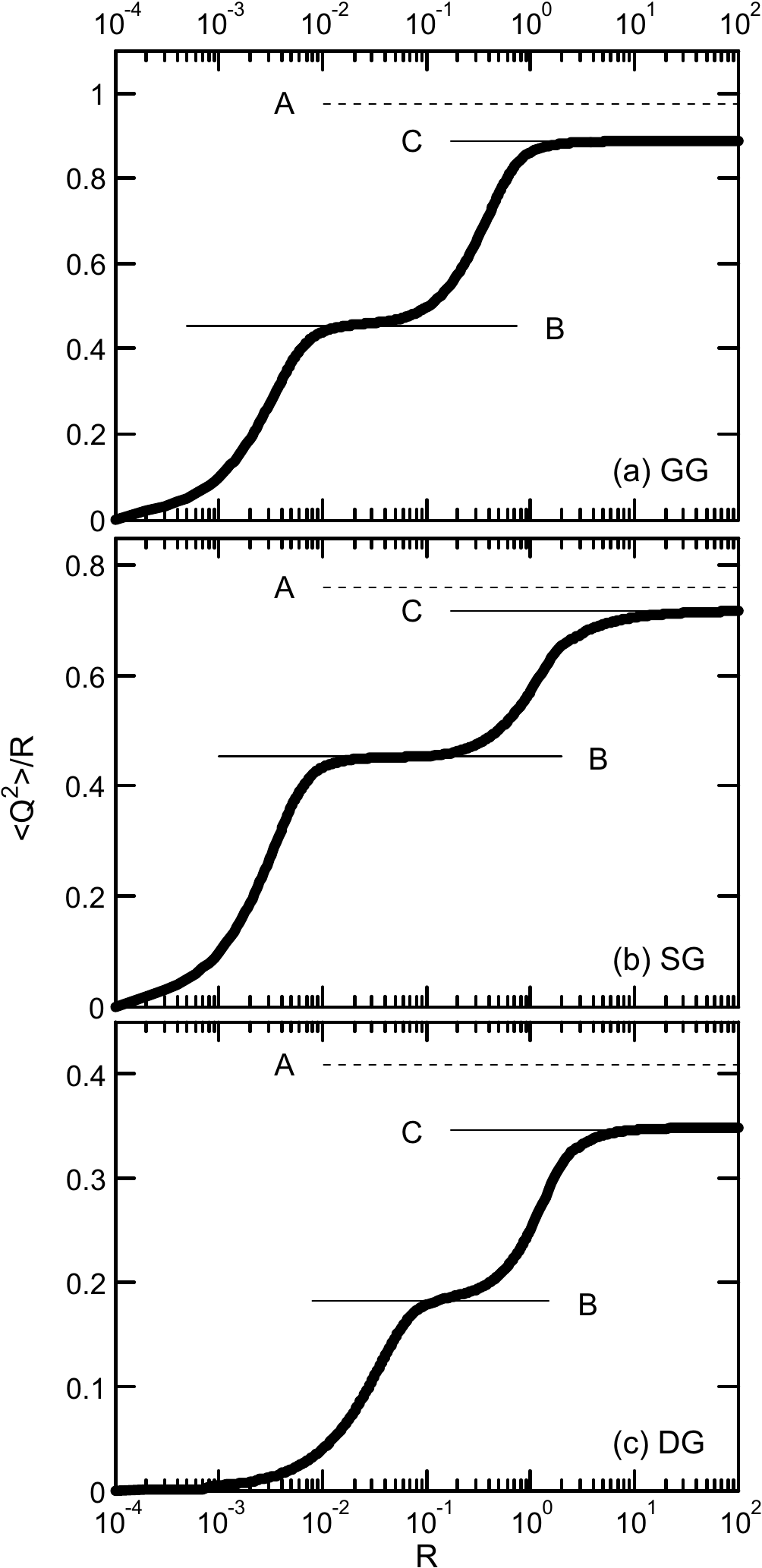}%
\caption{$\left\langle Q^{2}\right\rangle /R$ vs. $R$ for composite sources in
which both the $a$ and $b$ beams have short range correlations. \ The thick
curves are the exact result in Eq. (\ref{Q2David}). \ The solid lines labeled
B are the theory in Eqs. (\ref{SSQ2b})- (\ref{SSQ2bS}), the dashed lines
labeled A are the theory in Eqs. (\ref{SSQaPlusQb})- (\ref{SSQ2aSinc}),
whereas the solid lines labeled C are the short range screening result, Eq.
(\ref{Q2short}). \ (a) GG, beams $a$ and $b$ are Gaussians. \ Beam parameters
are $a=1,b=100,K=0.01$. \ (b) SG, beam $a$ is a Sinc, beam $b$ is a Gaussian.
\ Beam parameters are the same as in (a). \ \ (c) DG, beam $a$ is a Disk, beam
$b$ is a Gaussian. \ Beam parameters are $a=1,b=10,K=0.04$. \ In all three
examples displayed here the $b$ field was chosen to be a Gaussian in order to
emphasize that the agreement with the theory represented by line B does not
depend on the nature of the $a$ beam. \ Other short range choices for the $b$
beam, Disk or Sinc, show equally good agreement.}%
\label{Fig3}%
\end{center}
\end{figure}

We illustrate the above in Fig. \ref{Fig3} . \ As can be seen, Eqs.
(\ref{SSQ2bG})-(\ref{SSQ2bS}) are good approximations to the exact results,
whereas Eqs. (\ref{SSQaPlusQb})-(\ref{SSQ2aSinc}) are good only to order of
$10\%$.

We note that the well defined steps in this figure provide striking visual
confirmation of the fact that there are two widely different length scales.
\ The first step starts, as expected, at $R\sim1/b$, the second at $R\sim1/a$.

\section{BEAM $a$ EXHIBITS LONG-RANGE SCREENING, BEAM $b$ EXHIBITS SHORT-RANGE
SCREENING}

As discussed in the previous section, in region II, $1/b<R<1/a$, $\left\langle
Q^{2}\right\rangle $ is dominated by $\left\langle Q^{2}\right\rangle _{b}$,
Eqs. (\ref{SSQ2bG})- (\ref{SSQ2bS}). \ In region III, where $R>1/a$, beam $a$
exhibits long-range screening, and $\left\langle Q^{2}\right\rangle _{a}$ is
given by Eq. (\ref{LongRangeQ2}) with $p=a$. \ Neglecting again cross
screening of $a$ and $b$ charges, the total charge variance in region III is
approximated by the sum of $a$ and $b$ beam contributions, Eq.
(\ref{SSQaPlusQb}).

\begin{figure}
[pb]
\begin{center}
\includegraphics[
natheight=5.686100in,
natwidth=3.986800in,
height=3.4394in,
width=2.4189in
]%
{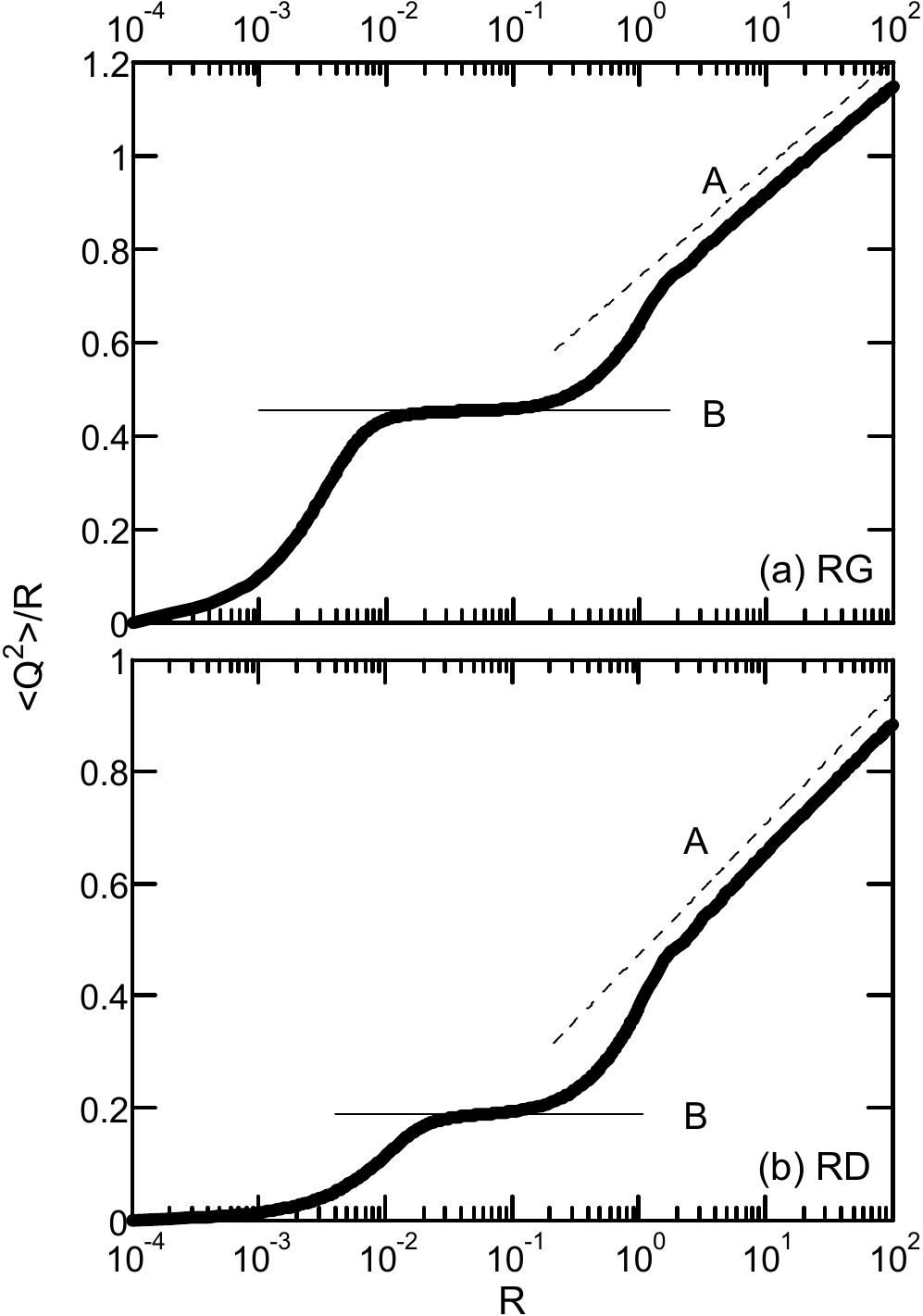}%
\caption{$\left\langle Q^{2}\right\rangle /R$ vs. $R$ for composite sources in
which the $a$ beam has long-range and the $b$ beam short-range correlations.
\ The thick curves are the exact result in Eq. (\ref{Q2David}). \ The solid
lines labeled B are the short-range screening theory in Eqs. (\ref{SSQ2b})-
(\ref{SSQ2bD}), the dashed lines labeled A are the long-range screening
result in Eq. (\ref{LongRangeQ2}). \ (a) RG, beam $a$ is a Ring, $b$ a
Gaussian. \ (b) RD, beam $a$ is a Ring, beam $b$ a Disk. \ Beam parameters in
both (a) and (b) are $\ a=1,b=100,K=0.01$.}%
\label{Fig4}%
\end{center}
\end{figure}

In Fig. \ref{Fig4} we plot $\left\langle Q^{2}\right\rangle /R$ vs. $R$ for RG
and DG, again obtaining in the region of short range screening, region B, a
plateau that is in good agreement with the calculated value for $\left\langle
Q^{2}\right\rangle _{b}$. \ As expected, in region A, the region of long range
screening, $\left\langle Q^{2}\right\rangle /R$ grows linearly with $\ln R$.
\ We note that here not only is the calculated slope, Eq. (\ref{LongRangeQ2}),
in close agreement with the exact result (thick curves), but also that Eq.
(\ref{SSQaPlusQb}) provides a rather reasonable description of the data.
\ Similar good agreement is obtained for RS (not shown).%

\section{BEAM $a$ EXHIBITS SHORT-RANGE SCREENING, BEAM $b$ EXHIBITS LONG-RANGE
SCREENING}

In region II, $1/b<R<1/a$, we again set $a=0$ in $W^{\left(  \text{T}%
_{a}\text{T}_{b}\right)  }$, and obtain for beam $b$ a Ring,%
\begin{equation}
\left\langle Q^{2}\right\rangle _{b}\approx\frac{Kb^{2}}{4\pi}\int_{0}%
^{2R}\sqrt{4R^{2}-r^{2}}\frac{J_{1}^{2}\left(  br\right)  }{1-J_{0}\left(
br\right)  }dr. \label{Q2bLR}%
\end{equation}
As before, we have assumed $K\ll1$, $\left[  KW_b^{\left(  \text{T}\right)
}\right]  ^{2}\ll W_b^{\left(  \text{T}\right)  }$. \ Proceeding as in
[$19$, (Eqs. 40-48)], we obtain
\begin{subequations}
\label{myQ2b}%
\begin{align}
\left\langle Q^{2}\right\rangle _{b}  &  \approx\dfrac{KbR}{2\pi^{2}}\left[
{\cal{F}}+\ln(bR)  \right]  ,\label{myQ2bLine1}\\
{\cal{F}}  &  =\pi {\cal{I}}+\gamma+5\ln2-3\approx4.84258, \label{myQ2Line2}%
\\
{\cal{I}}  &  =\int_{0}^{\infty}\frac{J_{0}\left(  x\right)  J_{1}^{2}\left(
x\right)  }{1-J_{0}\left(  x\right)  }dx\approx1.20946. \label{myQ2bLine3}%
\end{align}

We illustrate the above in Fig. \ref{Fig5}. \ In region B, $10^{-2}<R<10^{-1}%
$, for small $K$ ($0.015$) Eqs. (\ref{myQ2b}) provide a good description of
the exact result, Fig. \ref{Fig5}(a), whereas when $K$ is no longer small
($0.2$) the expected deviations appear, Fig. \ref{Fig5}(b).%
\begin{figure}
[pb]
\begin{center}
\includegraphics[
natheight=5.783900in,
natwidth=3.944400in,
height=3.4973in,
width=2.3938in
]%
{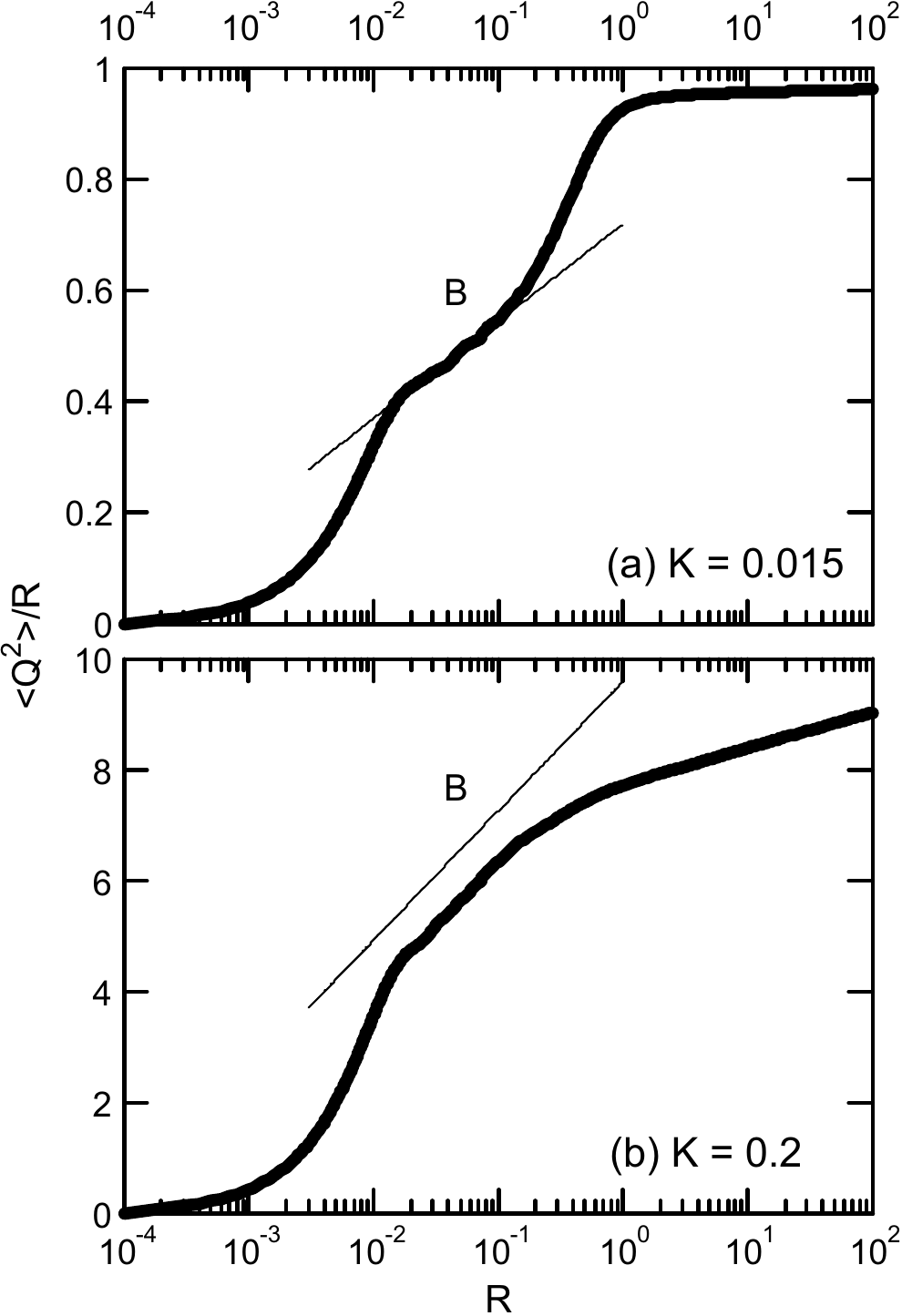}%
\caption{$\left\langle Q^{2}\right\rangle /R$ vs. $R$ for composite sources in
which beam $a$\ is a Gaussian with short-range correlations, and beam $b$ a
Ring with long-range correlations. \ Beam parameters are $\ a=1,b=100$. \ The
thick curves are the exact result in Eq. (\ref{Q2David}). \ The solid lines
labeled B are the long-range screening theory in Eqs. (\ref{myQ2b}). \ (a)
$K=0.015$. \ (b) $K=0.2.$}%
\label{Fig5}%
\end{center}
\end{figure}

In Fig. \ref{Fig5}(a) a plateau appears for $R>1$, apparently consistent with
the fact that short-range screening $a$ field charges become important in this
region. \ But this apparent plateau is misleading, because the long-range
screening of the $b$ field charges can never saturate: regardless of the
nature of the $a$ field, $\left\langle Q^{2}\right\rangle /R$ for the $b$
charges must diverge logarithmically for large $R$. \ The rate (slope) of this
logarithmic divergence, however, is $K$ dependent, being small, Fig.
\ref{Fig5}(a) (large, Fig. \ref{Fig5}(b)) for small (large) $K$.

\section{BOTH BEAM $a$ AND BEAM $b$ EXHIBIT LONG-RANGE SCREENING}

The case of two Rings is illustrated in Fig. \ref{Fig6}. \ As before, Eq.
(\ref{myQ2b}) holds for the $b$ field charges. \ But now the $a$ field charges
also exhibit long-range screening, and $\left\langle Q^{2}\right\rangle /R$
exhibits the expected large $R$ logarithmic divergence also for small $K$
($0.015$).%

\begin{figure}
[h]
\begin{center}
\includegraphics[
natheight=2.867700in,
natwidth=3.928000in,
height=1.7495in,
width=2.3843in
]%
{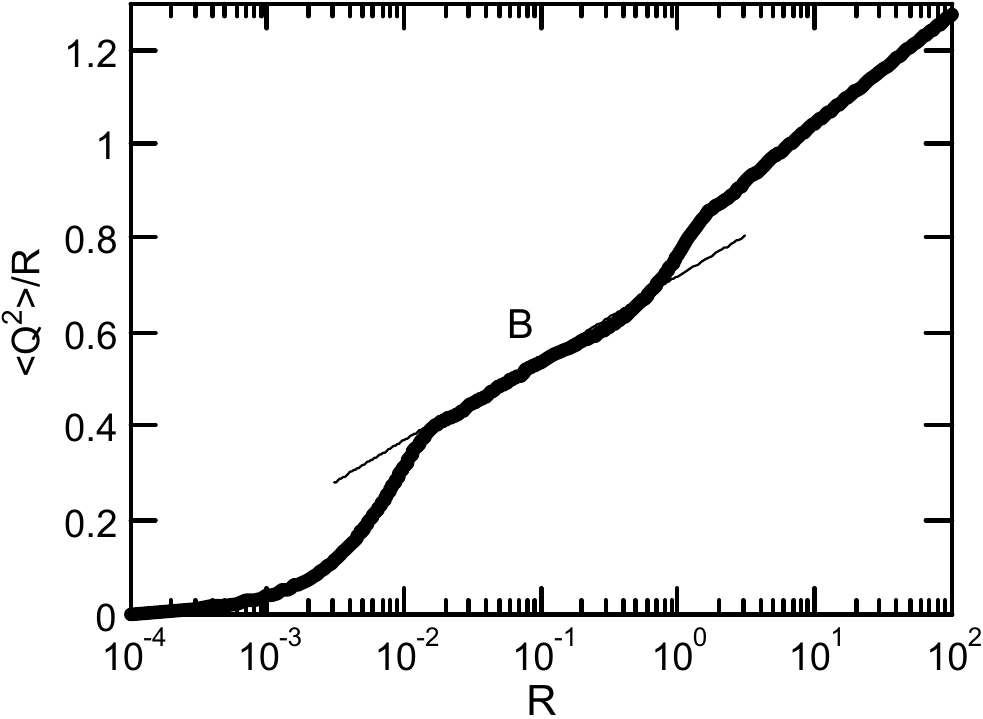}%
\caption{$\left\langle Q^{2}\right\rangle /R$ vs. $R$ for a composite source
in which both beam $a$\ and beam $b$ are Rings. \ Beam parameters are
$\ a=1,b=100$, $K=0.015$. \ The thick curve is the exact result in Eq.
(\ref{Q2David}). \ The solid line labeled B is the long-range screening
theory in Eqs. (\ref{myQ2b}).}%
\label{Fig6}%
\end{center}
\end{figure}

\section{SUMMARY}

Screening in speckled speckle produced by composite sources has been studied
in the limit of two widely different characteristic length scales in the
speckle pattern and two widely different intensities in the sources. \ The
interesting combination, emphasized here, is an intense field with a long
length scale (the $a$ field) perturbed by a weak field that has a short length
scale (the $b$ field).

Exact, and approximate, results have been presented for all four combinations
of short- and long-range screening. \ When the correlations in both fields are
short-range, cross screening between $a$ and $b$ field singularities appears
to be of only secondary importance, and the results of the exact calculation
are found to decompose, approximately, into a sum of screening contributions,
one for each field, Fig. \ref{Fig3}. \ A similar decomposition is found to
hold if the $b$ field exhibits short-range screening and the $a$ field
exhibits long-range screening, Fig. \ref{Fig4}. \ If the $b$ field exhibits
long-range screening, however, then no decomposition is valid, Figs.
\ref{Fig5} and \ref{Fig6}.

The unusual screening properties of speckled speckle are most pronounced when
the ratio ($b$ field to $a$ field) of length scales is $100:1$ or greater, and
the ratio of optical powers in the sources (the parameter $K$) is $1:100$ or
less. \ Less extreme parameter ratios can also yield useful results, however,
as in Fig. \ref{Fig3}(b), so that the experimental study of screening in speckled
speckle appears to be entirely feasible.
\end{subequations}
\begin{acknowledgments}
D. A. Kessler acknowledges the support of the Israel Science Foundation.
\end{acknowledgments}

\begin{center}
{\large \textbf{References}}
\end{center}

\hspace{-0.15in}[1] J. F. Nye and M. V. Berry, \textquotedblleft Dislocations
in wave trains,\textquotedblright\ Proc. Roy. Soc. London Ser. A \textbf{336,}
2165 (1974). For additional sources see online citation databases for the
numerous papers that reference this work.

\hspace{-0.15in}[2] J. F. Nye, \emph{Natural Focusing and the Fine Structure
of Light} (IOP Publ., London, 1999).

\hspace{-0.15in}[3] J. W. Goodman, \emph{Speckle Phenomena In Optics} (Roberts
\& Co., Englewood, Colorado, 2007).

\hspace{-0.15in}[4] B. I. Halperin, \textquotedblleft Statistical mechanics of
topological defects,\textquotedblright\ in \emph{Physics of Defects}, R.
Balian, M. Kleman, and J.-P. Poirier, eds. (North-Holland, Amsterdam, 1981),
p. 814-857.

\hspace{-0.15in}[5] F. Liu and G. F. Mazenko, \textquotedblleft Defect-defect
correlation in the dynamics of first-order phase
transitions,\textquotedblright\ Phys. Rev. B \textbf{46,} 5963-5971 (1992).

\hspace{-0.15in}[6] B. W. Roberts, E, Bodenschatz, and J. P. Sethna,
\textquotedblleft A bound on the decay of defect-defect correlation functions
in two- dimensional complex order parameter equations,\textquotedblright%
\ Physica D \textbf{99,} 252-268 (1996).

\hspace{-0.15in}[7] I. Freund and M. Wilkinson, \textquotedblleft
Critical-point screening in random wave fields,\textquotedblright\ J. Opt.
Soc. Am. A \textbf{15,} 2892-2902 (1998).

\hspace{-0.15in}[8] M. V. Berry and M. R. Dennis, \textquotedblleft Phase
singularities in isotropic random waves,\textquotedblright\ Proc. Roy. Soc.
London A \textbf{456,} 2059-2079 (2000); ibid., p. 3048.

\hspace{-0.15in}[9] M. V. Berry and M. R. Dennis, \textquotedblleft
Polarization singularities in isotropic random vector waves,\textquotedblright%
\ Proc. Roy. Soc. Lond. A \textbf{457,} 141-155 (2001).

\hspace{-0.15in}[10] M. R. Dennis, \textquotedblleft Polarization
singularities in paraxial vector fields: morphology and
statistics,\textquotedblright\ Opt. Commun. \textbf{213,} 201-221 (2002).

\hspace{-0.15in}[11] I. Freund, M. S. Soskin, and A. I. Mokhun,
\textquotedblleft Elliptic critical points in paraxial
fields,\textquotedblright\ Opt. Commun. \textbf{208,} 223-253 (2002).

\hspace{-0.15in}[12] G. Foltin, \textquotedblleft Signed zeros of Gaussian
vector fields - density, correlation functions, and
curvature,\textquotedblright\ J. Phys. A: Math. Gen. \textbf{36, }1729-1742 (2003).

\hspace{-0.15in}[13] M. R. Dennis, \textquotedblleft Correlations and
screening of topological charges in Gaussian random fields,\textquotedblright%
\ J. Phys. A: Math. Gen. \textbf{36,} 6611-6628 (2003).

\hspace{-0.15in}[14] M. Wilkinson, \textquotedblleft Screening of charged
singularities of random fields,\textquotedblright\ J. Phys. A: Math. Gen.
\textbf{37,} 6763-6771 (2004).

\hspace{-0.15in}[15] G. Foltin, S. Gnutzmann, and U. Smilansky,
\textquotedblleft The morphology of nodal lines - random waves versus
percolation,\textquotedblright\ J. Phys. A: Math. Gen. \textbf{37,}
11363-11372 (2004).

\hspace{-0.15in}[16] B. A. van Tiggelen, D. Anache, and A. Ghysels,
\textquotedblleft Role of mean free path in spatial phase correlation and
nodal screening,\textquotedblright\ Europhys. Lett. \textbf{74,} 999-1005 (2006).

\hspace{-0.15in}[17] I. Freund, R. I. Egorov, and M. S. Soskin,
\textquotedblleft Umbilic point screening in random optical
fields,\textquotedblright\ Opt. Lett. \textbf{22,} 2182-2184 (2007).

\hspace{-0.15in}[18] R. I. Egorov, M. S. Soskin, D. A. Kessler, and I. Freund,
\textquotedblleft Experimental measurements of topological singularity
screening in random paraxial scalar and vector optical
fields,\textquotedblright\ Phys. Rev. Lett. \textbf{100,} 103901 (2008).

\hspace{-0.15in}[19] D. A. Kessler and I. Freund, \textquotedblleft Short- and
long-range screening of optical phase singularities and C
points,\textquotedblright\ Opt. Commun. (2008), doi:10:1016/j.optcom.2008.05.018.

\hspace{-0.15in}[20] I. Freund and D. A. Kessler, \textquotedblleft
Singularities in speckled speckle,\textquotedblright\ Opt. Lett. \textbf{33,}
479-481 (2008).

\hspace{-0.15in}[21] I. Freund and D. A. Kessler, \textquotedblleft
Singularities in speckled speckle: Statistics,\textquotedblright%
\ arXive:0806365; Opt. Commun. (submitted).

\hspace{-0.15in}[22] J. W. Goodman, \emph{Statistical Optics} (John Wiley, New
York, 1985).

\hspace{-0.15in}[23] M. Berry, \textquotedblleft Disruption of wave-fronts:
statistics of dislocations in incoherent Gaussian random
waves,\textquotedblright\ J. Phys. \textbf{A 11,} 27-37 (1978).

\end{document}